\documentstyle[pramana,bm,epsf,epsfig,psfig,floats]{ias}
\begin{document}
\def\mpT{p_T \hspace{-1em}/\;\:}
\def\lsim    {\:\raisebox{-0.5ex}{$\stackrel{\textstyle<}{\sim}$}\:}
\def\gsim    {\:\raisebox{-0.5ex}{$\stackrel{\textstyle>}{\sim}$}\:}
\def\mev     {\: \rm MeV}
\def\gev     {\: \rm GeV}
\def\tot     {\: \rm tot}
\def\pp      {\: \rm pp}
\def\sigtot  {\mbox{$\sigma_{\rm tot}^{pp}$}}
\def\rs{\mbox{$\sqrt{s}$}}
\def\pbarp{\mbox{$ \bar{p}p$}}
\newcommand{\ra}{\rightarrow}
\newcommand{\ba}{\begin{array}}
\newcommand{\ea}{\end{array}}
\newcommand{\beqa}{\begin{eqnarray}}
\newcommand{\eeqa}{\end{eqnarray}}
\newcommand{\be}{\begin{equation}}
\newcommand{\ee}{\end{equation}}
\newcommand{\comment}[1]{}

\mark{{Theoretical expectations  for $\sigma^{tot}$ at the
LHC}{Rohini M Godbole et al}}

\title{Theoretical expectations  for \bm{$\sigma^{\rm tot}$} at the\\
large hadron collider}

\author{ROHINI M GODBOLE$^{1}$, AGNES GRAU$^{2}$, ROHIT
HEGDE$^{1}$,\\
GIULIA  PANCHERI$^{3}$ and YOGI SRIVASTAVA$^{4}$}
\address{$^1$Centre for High Energy Physics,
               Indian Institute of Science, Bangalore 560 012, India\\
$^2$Departamento de Fisica Teorica y del Cosmos,
               Universidad de Granada, Spain         \\
$^3$INFN, LNF, P.O. Box 13, I-00044 Frascati, Italy\\
$^4$Physics Department and INFN, University of Perugia,
            Perugia, Italy}

\abstract{
In this note, we summarize and compare various model predictions for $pp$
total cross-section $\sigma_{\rm tot}^{pp}$, giving an estimate of the
range of predictions for the total cross-section, $\sigma_{\rm tot}^{pp}$
expected at the LHC. We concentrate on the  results for $\sigma_{\tot}^{pp}$
obtained in a particular QCD based model of the energy dependence of the
total cross-section, including  the effect of soft gluon radiation.  We obtain
the range of predictions in this model by exploring the allowed range of model
parameters. We further give a handy parametrisation of these results which
incidentally spans  the range of various other  available predictions at the
LHC as well.}


\maketitle

\section{Introduction}

Energy dependence of total hadronic cross-sections has been the focus
of intense theoretical interest as a sensitive probe of strong
interactions long before the establishment of QCD as `the' theory of
hadrons. Even now, notwithstanding creditable successes of
perturbative and lattice QCD,  a first principle description of
total/elastic and inelastic hadronic cross-section is unavailable.
More pragmatically, for a correct projection of the expected
underlying activity at LHC, a reliable prediction of total
non-diffractive cross-section is essential to ensure the extraction
of new physics from the LHC data. Surely we will have to depend -- at
the initial stages of LHC -- upon predictions based on our current
understanding of these matters. Only much later it may become
feasible to use the LHC data itself towards this goal. Hence, a
critical evaluation of the range of theoretical predictions, is
absolutely essential.

The hadronic cross-section data exhibit, and require explanation of,
three basic features:
\pagebreak

\begin{description}
\itemsep-2pt
\item (i) the normalization of the cross section,
\item (ii)
an initial decrease and
\item (iii) a  subsequent rise with energy.
\end{description}
Various theoretical models exist which are  motivated by our
theoretical understanding of the strong interactions. The parameters
in these models, in most cases, are fitted to explain the observed
low energy data and the model predictions are then extrapolated to
give the \sigtot\ at the LHC energies. There are  different classes
of models. The highly successful Donnachie--Landshoff (DL)
parametrisation~\cite{Donnachie:1992ny} of the form
\begin{equation}
\sigma_{\rm tot}(s)=Xs^\epsilon+Ys^{-\eta}, \label{DL}
\end{equation}
has been used for a very long time. Here the two terms are understood
as arising from the Regge and the Pomeron trajectories, the
$\epsilon$ being approximately close to zero and $\eta$ being close
to $0.5$. These values seem to be consistent with a large, but not
all, body of the hadronic cross-sections. In this note we will  first
present  phenomenological arguments for the approximate values of
these parameters which seem to be required to describe the data
satisfactorily.  As a matter of fact, there also exist in the
literature discussions of the `hard' Pomeron~\cite{DLnew} motivated
by the discrepancies in the rate of energy rise observed by
E710~\cite{Amos:1991bp}, E811~\cite{Avila:1998ej} and the
CDF~\cite{Abe:1993xy}.  In addition, a variety of models exist
wherein the observed energy dependence of the cross-section, along
with few very general requirements of factorisation, unitarity and/or
ideas of finite energy sum rules (FESR), is used to determine the
values of model
parameters~\cite{Block:2005ka,Block:2005pt,Igi:2005jm,Avila:2002tk,Cudell:2002xe,Cudell:2002sy,Avila:2006ya}.
The so-obtained parametrisations are then extended to make
predictions at the LHC energies.  There also exist QCD motivated
models based on the  mini-jet
formalism~\cite{Cline:1973kv,Pancheri:1985sr,Gaisser:1984pg}, wherein
the energy rise of the total cross-sections is driven by the
increasing number of the low-$x$ gluon--gluon collisions. These
models need to be embedded in an eikonal
formalism~\cite{Durand:1987yv} to soften the violent energy  rise of
the mini-jet cross-sections. Even after eikonalisation the  predicted
energy rise is harder than the gentle one observed
experimentally~\cite{Gaisser:1984pg,Pancheri:1986qg}. A QCD based
model where the rise is further tamed by the  phenomenon of
increasing emission of soft gluons by the valence quarks in the
colliding hadrons, with increasing
energy~\cite{Grau:1999em,Godbole:2004kx}, offers a consistent
description of \sigtot.  Thus we have a variety of model predictions
for \sigtot\ at the LHC.  In this note we compare these predictions
with each other to obtain an estimate of the `theoretical'
uncertainty in them.

\section{Phenomenological models}

The two terms of eq. (\ref{DL})~\cite{Donnachie:1992ny} reflect the
well-known duality between resonance and Regge pole exchange on the
one hand and background and Pomeron exchange on the other,
established in the late 60's through FESR \cite{Igi:1967}. This
correspondence  meant that, while at low energy the cross-section
could be written as due to a  background term and a sum of
resonances,  at higher energy it could be written as a sum of Regge
trajectory exchanges and  a Pomeron exchange.

It is well to ask (i) where the  `two-component' structure of eq.
(\ref{DL}) comes from and (ii) why the difference in the two powers
(in $s$) is approximately a half. Our present knowledge of QCD and
its employment for a description of hadronic phenomena can be used to
provide some insight into the `two-component' structure of
eq.~(\ref{DL}). This begins with considerations about the bound state
nature of hadrons which necessarily transcends perturbative QCD. For
hadrons made of light quarks ($q$) and gluons ($g$), the two terms arise
from $q\bar{q}$ and $gg$ excitations. For these, the energy is given
by a sum of three terms: (i) the rotational energy, (ii) the Coulomb
energy and (iii) the `confining' energy. If we accept the Wilson area
conjecture in QCD, (iii) reduces to the linear
potential~\cite{Srivastava:2000fb,Landshoff:2001pp}. Then the
hadronic rest mass for a state of angular momentum $J$ can be
obtained by minimising  the expression for the energy of two massless
particles ($q \bar q$ or $gg$) separated by a distance $r$.

Explicitly, in the CM frame of two massless particles,
either a $q\bar{q}$ or a $gg$ pair separated by a relative
distance $r$ with relative angular momentum $J$, the energy is given by
\begin{equation} \label{string1}
E_i(J, r) = {{2J}\over{r}} - {{C_i \bar{\alpha}}\over{r}} + C_i \tau r,
\end{equation}
where $i = 1$ refers to $q\bar{q}$,  $i = 2$ refers to $gg$, $\tau$
is the `string tension' and the Casimir's are $C_1 = C_{\rm F} =
4/3$, $C_2 = C_{\rm G} = 3$. $\bar{\alpha}$ is the QCD coupling
constant evaluated at some average value of $r$ and whose precise
value will disappear in the ratio to be considered. The hadronic rest
mass for a state of angular momentum $J$ is then computed through
minimising the above energy
\begin{equation} \label{string2}
M_i(J) = \min_r\left[{{2J}\over{r}} - {{C_i \bar{\alpha}}\over{r}} +
C_i \tau r \right].
\end{equation}
This is then given by
\begin{equation} \label{string3}
M_i(J) = 2 \sqrt{(C_i \tau)[2J - C_i \bar{\alpha}]}.
\end{equation}

This can then be used to obtain the two sets of linear Regge
trajectories
\begin{equation} \label{string4}
\alpha_i(s) = {{C_i \bar{\alpha}}\over{2}} + \left({{1}\over{8 C_i
\tau}}\right) s = \alpha_i(0) + \alpha_i' s,
\end{equation}
Note that $\alpha_i$
are {\it not} the coupling constants.

Thus, the ratio of the intercepts is given by
\begin{equation}\label{string5}
{{\alpha_{gg}(0)}\over{\alpha_{q\bar{q}}(0)}} = C_{\rm G}/C_{\rm F} =
{{9}\over{4}}.
\end{equation}

Employing our present understanding that resonances are $q{\bar q}$
bound states while the background, dual to the Pomeron, is provided
by gluon--gluon exchanges\cite{Landshoff:2001pp}, the above equation
can be rewritten  as
\begin{equation} \label{string5qhd}
{{\alpha_{\rm P}(0)}\over{\alpha_{\rm R}(0)}} = C_{\rm G}/C_{\rm F} =
{{9}\over{4}}.
\end{equation}
If we restrict our attention to the leading Regge trajectory, namely
the degenerate $\rho-\omega-\phi$ trajectory, then
 $\alpha_{\rm R}(0)=\eta \approx 0.48$--$0.5$, and we obtain
for $\epsilon \approx 0.08$--$0.12$, a rather  satisfactory value. The
same argument for the slopes gives
\begin{equation}\label{string6}
{{\alpha_{gg}'}\over{\alpha_{q\bar{q}}'}} = C_{\rm F}/C_{\rm G} = {{4}\over{9}}.
\end{equation}
Hence, if  we take for the Regge slope $\alpha_{\rm R}' \approx
0.88$--$0.90$, we get for $\alpha_{\rm P}' \approx 0.39$--$0.40$,  in
fair agreement with lattice estimates\cite{lattice:2005}.

We now have good reasons for a break up of the amplitude into two
components. To proceed further, it is necessary to realize that
precisely because massless hadrons do not exist,  eq. (\ref{DL})
violates the Froissart bound and thus must be unitarized. To begin
this task, let us first rewrite  eq. (\ref{DL}) by putting in the
`correct' dimensions
\begin{equation} \label{DL1}
\bar{\sigma}_{\rm tot}(s)= \sigma_1 (s/\bar{s})^\epsilon+ \sigma_2
(\bar{s}/s)^{1/2},
\end{equation}
where we have imposed the nominal value $\eta = 1/2$.  It is possible
to  obtain~\cite{Godbole:2004kx} rough estimates for the
size of the parameters in eq. (\ref{DL1}).  A minimum occurs in
$\bar{\sigma}_{\rm tot}(s)$ at $s = \bar{s}$, for $\sigma_2 =
2\epsilon \sigma_1$.  If we make this choice, then eq. (\ref{DL1})
has one less parameter and it reduces to
\begin{equation} \label{DL2}
\bar{\sigma}_{\rm tot}(s)= \sigma_1 [(s/\bar{s})^\epsilon+ 2 \epsilon
(\bar{s}/s)^{1/2}].
\end{equation}
We can isolate the rising part of the cross-section by rewriting the above
as
\begin{equation}\label{DL3}
\bar{\sigma}_{\rm tot}(s)= \sigma_1 [ 1 + 2\epsilon(\bar{s}/s)^{1/2}]
+ \sigma_1 [(s/\bar{s})^\epsilon - 1].
\end{equation}
Equation (\ref{DL3}) separates cleanly the cross-section into two
parts: the first part is a `soft' piece which shows a saturation to a
constant value (but which contains no rise) and the second a `hard'
piece which has all the rise.  Moreover, $\bar{s}$ naturally provides
the scale beyond which the cross-sections would begin to rise. Thus,
our `Born' term assumes the generic form
\begin{equation} \label{DL4}
\sigma_{\rm tot}^{\rm B}(s)= \sigma_{\rm soft}(s) + \vartheta (s -
\bar{s}) \sigma_{\rm hard}(s)
\end{equation}
with $\sigma_{\rm soft}$ containing a constant (the `old' Pomeron
with $\alpha_{\rm P}(0) = 1$) plus a (Regge) term decreasing as
$1/\sqrt{s}$ and with an estimate for their relative magnitudes
($\sigma_2/\sigma_1 \sim 2\epsilon$).
 In the eikonalised mini-jet model used
by us~\cite{Godbole:2004kx} the rising part of the cross-section
$\sigma_{\rm hard}$ is provided by jets which are calculable by
perturbative QCD, obviating (at least in principle) the need of an
arbitrary parameter $\epsilon$. An estimate of $\sigma_1$ can also be
obtained~\cite{Godbole:2004kx} and is $\sim 40$ mb.

As said earlier, the DL parametrisation~\cite{Donnachie:1992ny} is a
fit to the existing data of the form given by eq. (\ref{DL}), with
$\epsilon =0.0808, \eta = 0.4525$. This fit has been extended to
include a `hard' Pomeron~\cite{DLnew} due to the discrepancy between
different data sets. The BH model~\cite{Block:2005ka} gives a fit to
the data using duality constraints. The BH fit for $\sigma^{\pm} =
\sigma^{\bar p p}/ \sigma^{p p}$ as a function of beam energy $\nu$,
is given as
\myeqn{\sigma^{\pm} = c_0 +c_1 \ln(\nu/m) + c_2 \ln^2
(\nu /m) + \beta_{{\rm P}'} (\nu/m)^{\mu-1} \pm \delta (\nu/m)^{\alpha -1},
\nonumber}
\eqnreset
\noindent
where  $\mu = 0.5, \alpha = 0.415$ and all the other parameters in mb are
$c_0 = 37.32, c_1 = -1.440 \pm 0.07, c_2 = 0.2817 \pm 0.0064, \beta_{{\rm P}'} =
37.10, \delta = -28.56$.
The fit obtained by Igi and Ishida \cite{Igi:2005jm} using the finite energy sum
rules (FESR) gives LHC predictions very similar to those given by the BH [6] fit.
Avila {\it et al} \cite{Avila:2002tk} gave a fit using analyticity arguments
whereas Cudell {\it et al}~\cite{Cudell:2002xe} gave predictions at the LHC energies
by extrapolating fits obtained to the current data again using constraints
from unitarity, analyticity of the S-matrix, factorisation,
coupled with a requirement that the cross-section asymptotically goes to
a constant as $\ln (s)$ or  $\ln^ 2 (s)$, in the framework of the COMPETE program.

\begin{figure}[t!]
\hskip4pc
\includegraphics[scale=0.60]{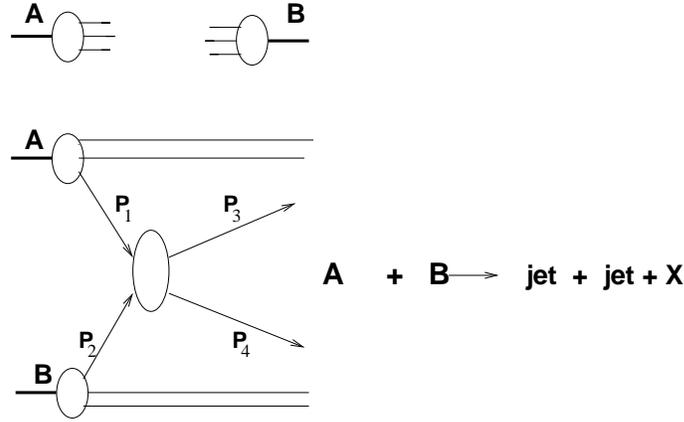}
\vspace{0.5em}

\caption{Parton model picture for jet production.}
\label{partonmodel}
\end{figure}

In the mini-jet models the energy rise of \sigtot\ is driven by the
increase with energy of the $\sigma_{\rm jet}^{AB}$ given by
\setcounter{equation}{12}
\beqa
~~~~~~~~ \sigma^{AB}_{\rm jet} (s)& = & \int_{p_{t\min}}^{\rs/2} {\rm
d} p_t \int_{4 p_t^2/s}^1 {\rm d} x_1 \int_{4 p_t^2/(x_1 s)}^1 {\rm d} x_2
\sum_{i,j,k,l} f_{i|A}(x_1) f_{j|B}(x_2)\nonumber \\
& & \times \frac {
{\rm d} \hat{\sigma}_{ij \rightarrow kl}(\hat{s})} {{\rm d} p_t},
\label{sigjet}
\eeqa
where
subscripts $A$ and $B$ denote particles ($\gamma,  p, \dots$), $i, j,
k,  l$ are partons and $x_1,x_2$ the fractions of the parent particle
momentum carried by the parton. $\hat{s} = x_1 x_2 s$  and $\hat{
\sigma}$ are hard partonic scattering cross-sections. This parton
model
used in the calculation is illustrated in figure~\ref{partonmodel}.
Note here that the experimentally measured parton densities in the proton and
the elements of perturbative QCD is all the input needed for the calculation of
$\sigma_{\rm jet}^{AB}$. The rate of rise with energy of this cross-section
is determined by $p_{t\min}$ and the low-$x$ behaviour of the parton densities.
As said before, the rise with energy of this cross-section is much steeper than
can be tolerated with the Froissart bound. Hence it
has to be embedded in an eikonal formulation given by
\begin{equation}
\sigma^{AB}_{\rm tot}=2\int {\rm d}^2{\vec b}[1-{\rm e}^{-\Im
m\,\chi^{AB}(b,s)}],
\label{stot}
\end{equation}
where ${\rm Re}\,\chi^{AB} (b,s) = 0$ and
$2\Im m~\chi^{AB}(b,s)=n^{AB}(b,s)$ is the average number of multiple
collisions which are Poisson distributed.  As outlined in eq. (\ref{DL3}) this
quantity too has contributions coming from soft and hard physics and can be
written as
\beqa
n^{AB}(b,s) &=&  n^{AB}_{\rm soft} +  n^{AB}_{\rm hard} \nonumber \\
& \simeq & A^{AB}_{\rm soft}(b)
\sigma^{AB}_{\rm soft}(s) +  A^{AB}_{\rm jet}(b) \sigma^{AB}_{\rm jet}(s).
\hspace{0.6in}
\label{nsplit}
\eeqa
In the second step the number $n(b,s)$  has been  assumed  to be factorisable
into an overlap function $A(b)$ and the cross-section $\sigma$.

\begin{figure}
\hskip4pc
\includegraphics[width=6.5cm]{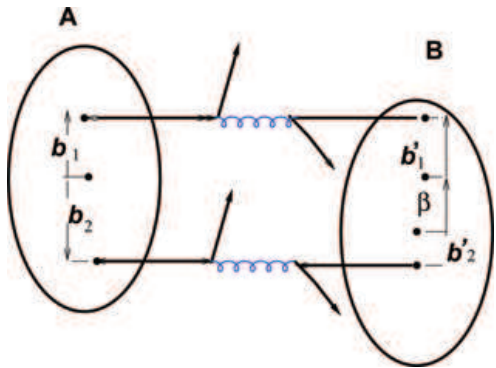}
\vspace{0.5em}

\caption{Multiple scattering giving rise to the
transverse overlap of the hadrons.}
\label{overlap}
\end{figure}

The sketch in figure 2 indicates the relationship between the
multiple scatterings and the overlap function.  The assumption
of factorisation as well as the split up between the two contributions, hard
and soft, for the $n(b,s)$ are only approximate. The extent to which this
softens the energy
rise, depends on the $b$ dependence of $n(b,s)$, i.e., that of $A(b)$ in the
factorised case. The normal assumption of using the same form of $A(b)$ for
both the hard and the soft part, given  by the Fourier transform of the
electromagnetic form factor (FF), still gives too steep a rise
even in this eikonalised mini-jet model (EMM)~\cite{Durand:1987yv}. In our model
this rise is tamed by including the effect on the transverse  momentum
distribution of the partons in the proton,  of the soft gluon emission from
the valence quarks in the proton~\cite{Grau:1999em}; the effect increasing with
increasing energy. In this description, the transverse momentum distribution
of the quarks in a proton can then be calculated in a semi-classical
picture as arising from the re-summation of a large number of soft gluons.
This in turn allows us to calculate the transverse overlap function.

\begin{figure}[t!]
\hskip4pc
\includegraphics[scale=0.65]{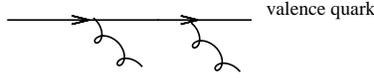}
\vspace{0.5em}

\caption{Multiple gluon emission giving rise to the transverse momentum
distribution of the valence quarks.}
\label{emission}
\end{figure}

Figure \ref{emission}  sketches the multiple emissions of gluons which
gives rise to the transverse momentum distribution of the valence
quarks and hence the overlap function.  The non-perturbative soft
part of the eikonal includes only limited low energy gluon emission
and leads to the initial decrease in the proton--proton
cross-section. On the other hand, the rapid rise in the hard, perturbative
jet part of the eikonal is tamed into the experimentally observed
mild increase by soft gluon radiation whose maximum energy
($q_{\max}$) rises slowly with energy. Thus the overlap functions
$A(b)$  are no longer a function of $b$ alone. We denote the
corresponding overlap function by $A_{BN} (b,q_{\max})$. The
\sigtot\ can then be computed using eq. (14) where $2\Im m\, \chi^{pp}(b,s)$ is
given by
\beqa
2\Im
m\,\chi(b,s)&=&n(b,s;q_{\max},p_{t\min})=n_{\rm soft} +
n_{\rm jet}\nonumber\\
&=&A_{\rm soft}(b,s)\sigma_{\rm soft} (s) +
A_{BN}(b,q_{\max})\sigma_{\rm jet} (s).
\eeqa

\noindent
The function $A_{BN}
(b,q_{\max})$~\cite{Grau:1999em} is determined by $q_{\max}$, which  in
turn depends on the energy and the kinematics of the subprocess. What
we use is its  average value over all the momentum fractions of the
parent partons. We need to further make a model for the `soft' part
$A_{\rm soft}$, which is determined by the non-perturbative dynamics. It
is this part of the eikonal that contributes to the \sigtot\ at high
energies through its impact on the turn around from the decreasing
Regge behaviour to the softly rising behaviour around $\sqrt{s}
\simeq 15$ GeV, where the  hard part contribution is minuscule.  Note
that we have taken $n_{\rm soft}$  to be factorised into a constant soft
cross-section $\sigma_{\rm soft}$ and taken $A_{\rm soft}$ to be given by the
function  $A_{BN}(b,q_{\max}^{\rm soft})$ except for the fact that it is
not possible to calculate the $q_{\max}^{\rm soft}$ for the soft
processes, as in the case of hard processes. We further postulate
that the $q_{\max}$ is the same for the hard and soft processes at low
energy, parting company around $10$ GeV where the hard processes
start becoming important. A good fit to the data requires
$q_{\max}$ at  low energies to be a very slowly decreasing function of
energy, with a value around 0.20 MeV at $\sqrt{s} = 5$ GeV rising
to about $0.24$ MeV, $\sqrt{s} \geq 10$ GeV, the upper value of this
soft scale being completely consistent with our picture of the
proton. Further, we need to fix one more parameter for
non-perturbative region, the $\sigma_{\rm soft}$. For the $pp$ case it is
a constant $\sigma_0$ which will fix the normalization of \sigtot,
whereas for the $p \bar p$ the duality arguments suggest that there
is an additional $\sqrt{s}$ dependent piece $\simeq 1/\sqrt{s}$. Thus
neglecting the real part of the eikonal, $n(b,s)$ in our model is
given by
\begin{equation}
n(b,s)  = A_{BN}(b,q_{\max}^{\rm soft}) \sigma_{\rm soft}^{pp,{\bar p}}+
A_{BN}(b,q_{\max}^{\rm jet})
 \sigma_{\rm jet}(s;p_{t\min}),
\end{equation}
where
\begin{equation}
\sigma_{\rm soft}^{pp}=\sigma_0,\quad \sigma_{\rm soft}^{p{\bar
p}}=\sigma_0 \left(1+{{2}\over{\sqrt{s}}}\right).
\end{equation}

\begin{figure}[t!]
       \centerline{
            \psfig{figure=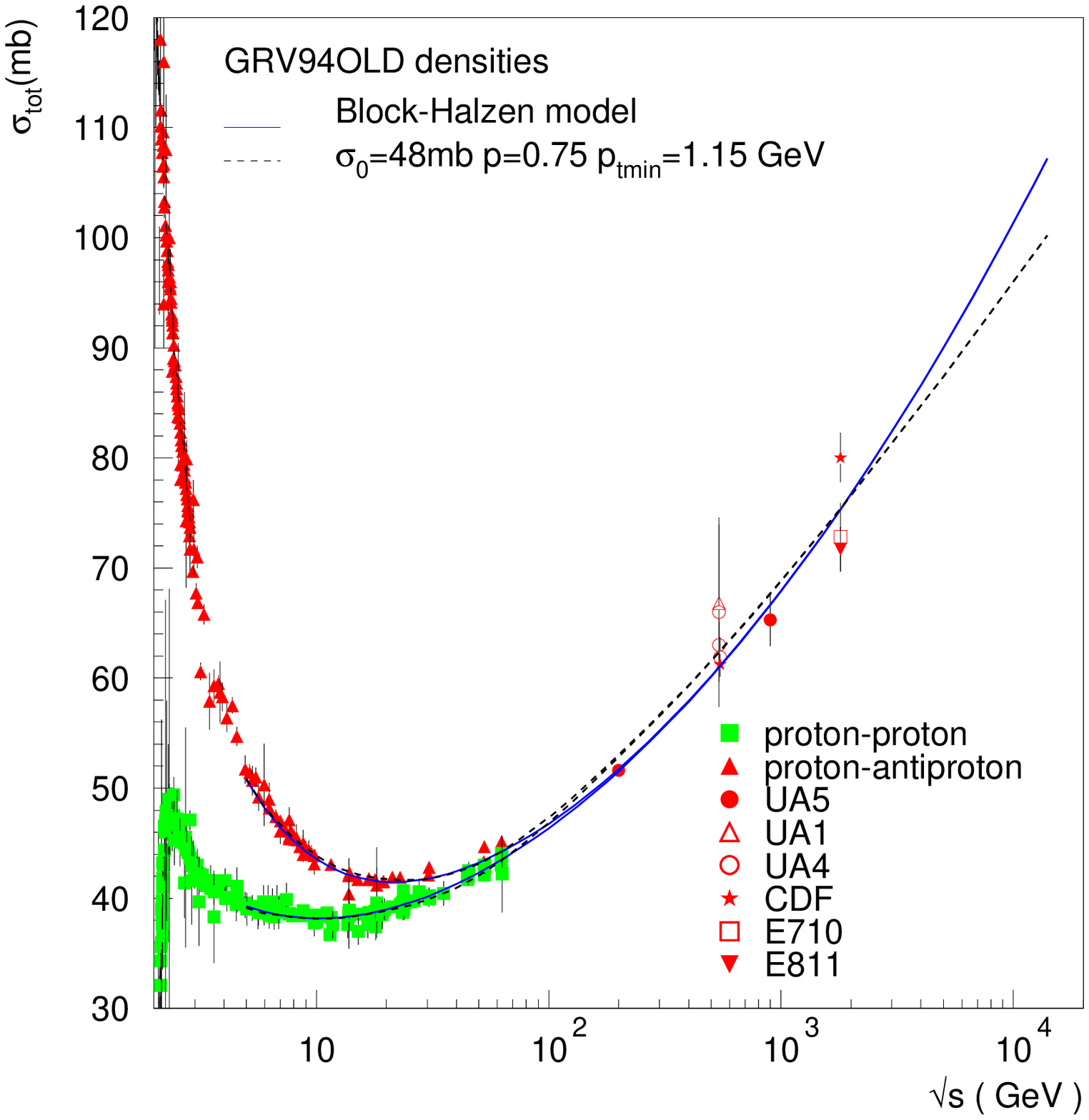,width=7cm,height=7cm,angle=0}
            \psfig{figure=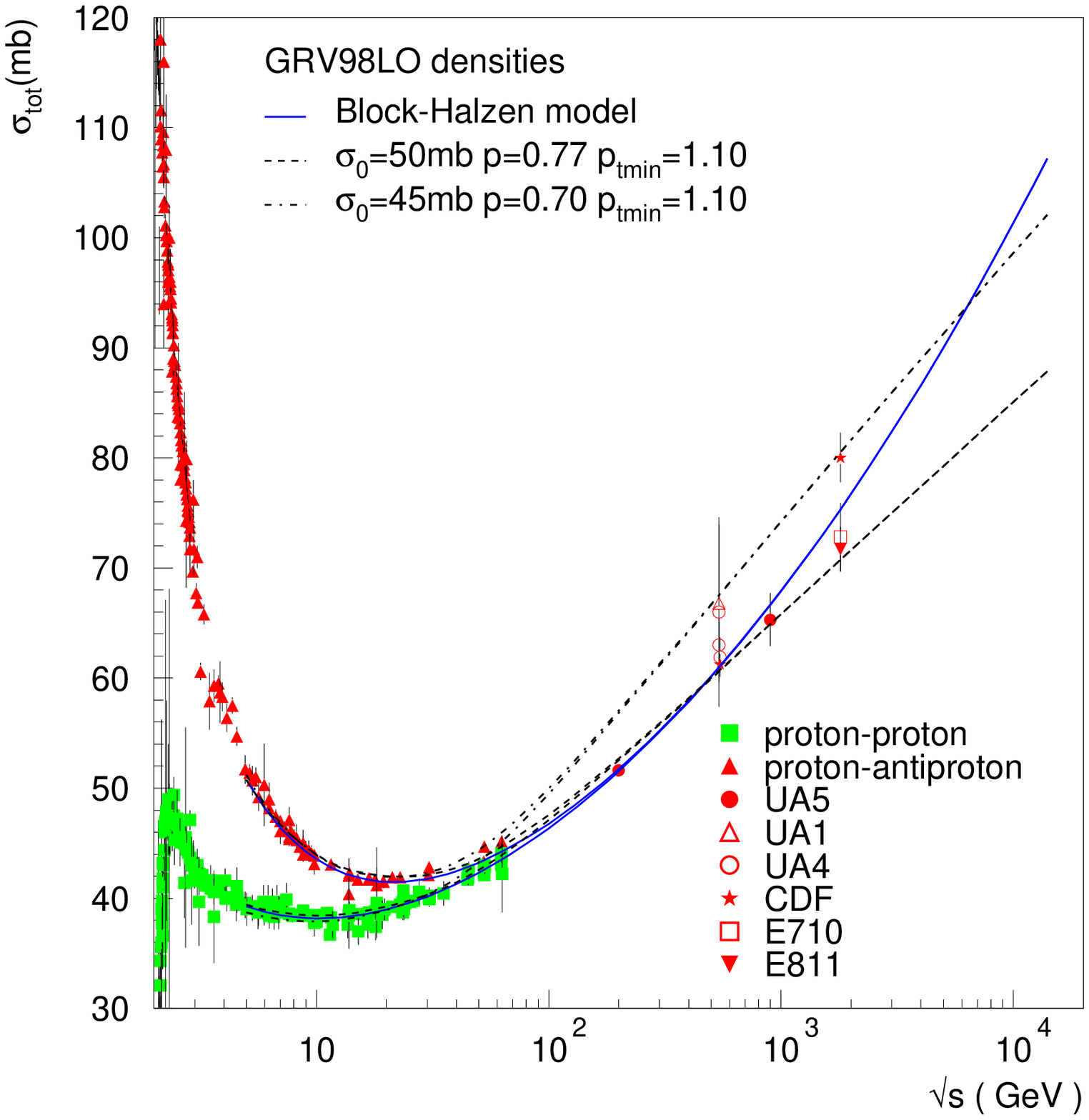,width=7cm,height=7cm,angle=0}}
\vspace{0.5em}

\caption{Comparison of the G.G.P.S. model \protect\cite{Godbole:2004kx}
predictions for GRV\protect\cite{Gluck:1991ng}  and
GRV98lo\protect\cite{Gluck:1998xa} densities with the
BH \protect\cite{Block:2005ka} predictions.  \label{bgrv}}
\end{figure}

\begin{figure}[t!]
\centerline{
            \psfig{figure=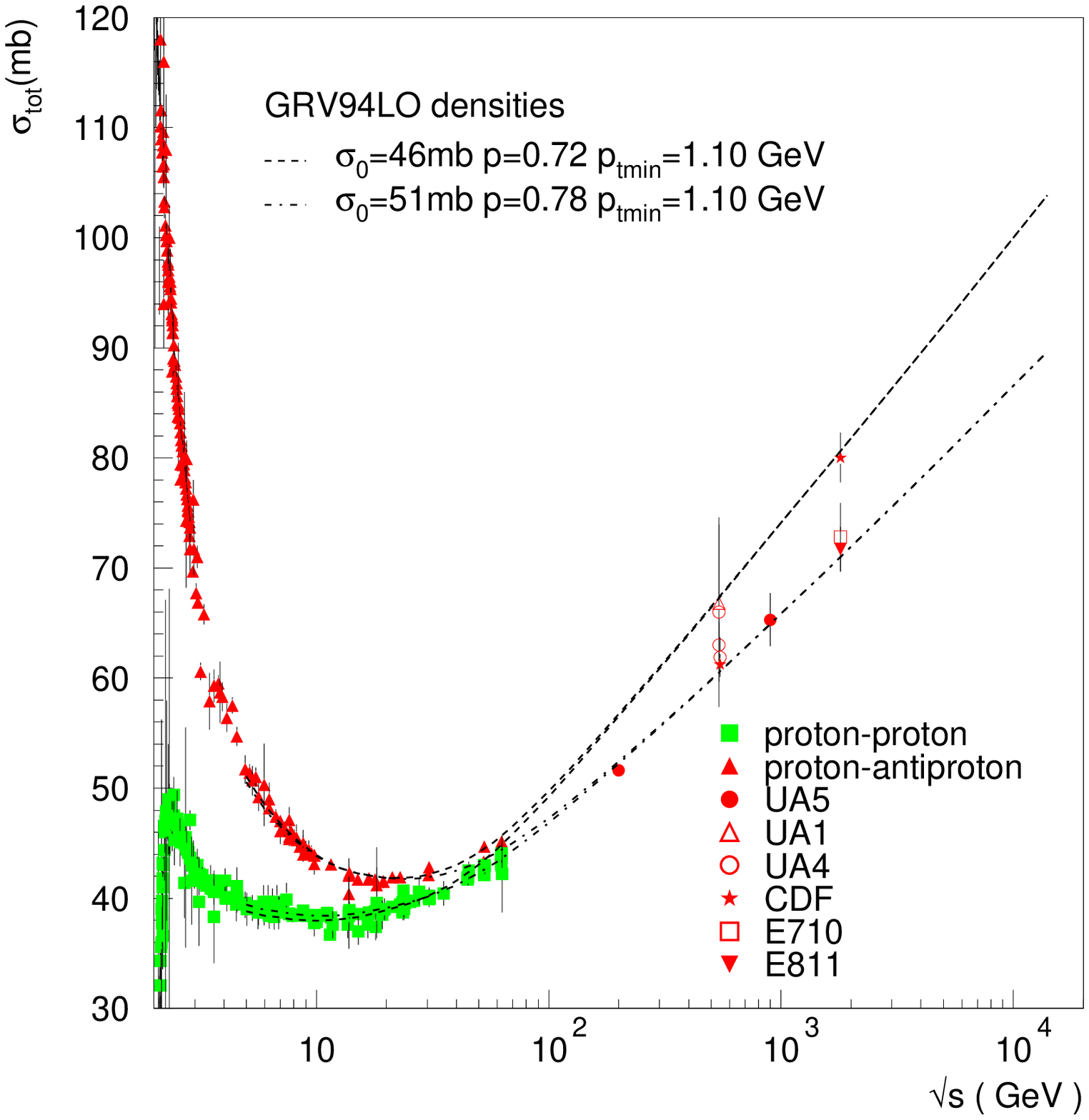,width=7cm,height=7cm,angle=0}
            \psfig{figure=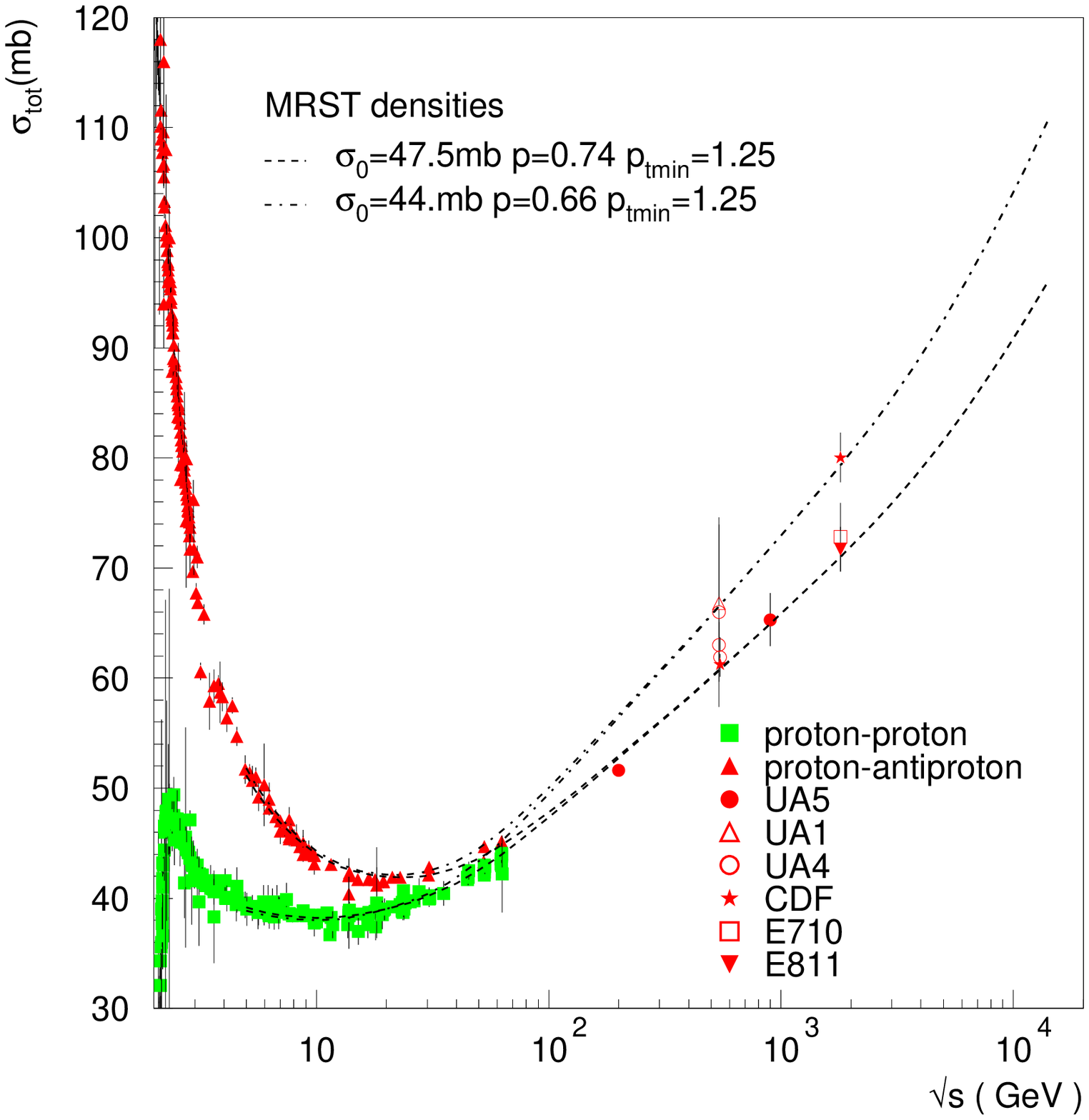,width=7cm,height=7cm,angle=0}}
\vspace{0.5em}

\caption{Predictions of the G.G.P.S model\protect\cite{Godbole:2004kx} for the
GRV94lo\protect\cite{Gluck:1994uf} and the
MRST~\protect\cite{Martin:2002dr} densities for model parameters
mentioned in table 1. \label{bgrvmrst}
}
\end{figure}

Thus the parameters of the model are $p_{t\min}$ and $\sigma_0$. In addition,
the evaluation of $A_{BN}$ involves $\alpha_s$ in the infrared region,
for which we use a phenomenological form inspired by the Richardson potential
\cite{Grau:1999em}. This involves a parameter $p$ which for the
Richardson potential takes value $1$.
Values of $p_{t\min}, \sigma_0$ and $p$  which give a good fit to the data
with the GRV parametrisation of the proton densities~\cite{Gluck:1991ng}
are  $1.15$ GeV,  $48$ mb  and $3/4$   respectively, as presented in
ref.~\cite{Godbole:2004kx}. These values are  consistent with the
expectations of the general argument~\cite{Godbole:2004kx}.  We expect these
best fit values to change somewhat with the choice of parton density functions
(PDF).  Since we are ultimately interested in the predictions of the model
at TeV energies, we need PDF parametrisation which cover both the small and
large $Q^2$ range ($2 < Q^2 < 10^4$) as well as are valid up to rather
small values of $x (\sim 10^{-5})$. Further, since our calculation here is only
LO, for consistency we have to use LO densities. We have repeated the exercise
then for a range of PDF's~\cite{Gluck:1994uf,Gluck:1998xa,Martin:2002dr}
meeting these requirements.  For each PDF, it is onset of the rise that
fixes the $p_{t\min}$, $\sigma_0$ controlling the normalisation and
$p$ determining the slope of the rising part of the cross-section.
We find that it is possible to get a satisfactory description of all the
current data, for all the choices of PDF's considered. The corresponding range
of values of $p_{t\min}, \sigma_0$ and $p$ are given in table 1.
Figure \ref{bgrv} compares with the data  the predictions of our QCD based model
along with  those obtained by the considerations of unitarity and
factorisation~\cite{Block:2005ka}. As can be seen clearly, both
are able to describe the current data on total cross-sections equally
satisfactorily.
As can be seen from figures~\ref{bgrv} and \ref{bgrvmrst}, it is possible to
get equally satisfactory description of the data in our QCD based model, for
all the chosen PDF's, by tuning the soft parameters by a small amount.
In the next section we compare these model results with the predictions of
all the other models.

\begin{table}[b!]
\hskip4pc
\begin{minipage}[t]{20pc}
{\small \bf Table 1.} {\small Values of $p_{t\min}$ and $\sigma_0$ corresponding to the
different parton densities in the proton, for  which the EMM (as
described in ref.~\protect\cite{Godbole:2004kx}) gives a satisfactory
description of \sigtot.}
\end{minipage}
\vspace{0.5em}

\hskip4pc\vbox{
\columnwidth=20pc
\begin{tabular}{lccc}
\\[-2mm]
PDF & $p_{t\min}$ (GeV) & $\sigma_0$ (mb)& $p$\\[1mm]
\hline\\[-2mm]
GRV \cite{Gluck:1991ng}     & 1.15 & 48\phantom{.0} &  0.75 \\[2mm]
GRV94lo \cite{Gluck:1994uf} & 1.10 & 46\phantom{.0} & 0.72 \\
                            & 1.10 & 51\phantom{.0} & 0.78 \\[2mm]
GRV98lo \cite{Gluck:1998xa} & 1.10 & 45\phantom{.0} & 0.70 \\
                            & 1.10 & 50\phantom{.0} & 0.77 \\[2mm]
MRST \cite{Martin:2002dr}   & 1.25 & 47.5 & 0.74 \\
                            & 1.25 & 44\phantom{.0} & 0.66\\[1mm]
\end{tabular}}
\end{table}
\setcounter{table}{1}

\section{Model predictions for \bm{\sigtot}\ at the LHC}

\begin{figure}[t!]
\hskip4pc
\includegraphics[scale=0.65]{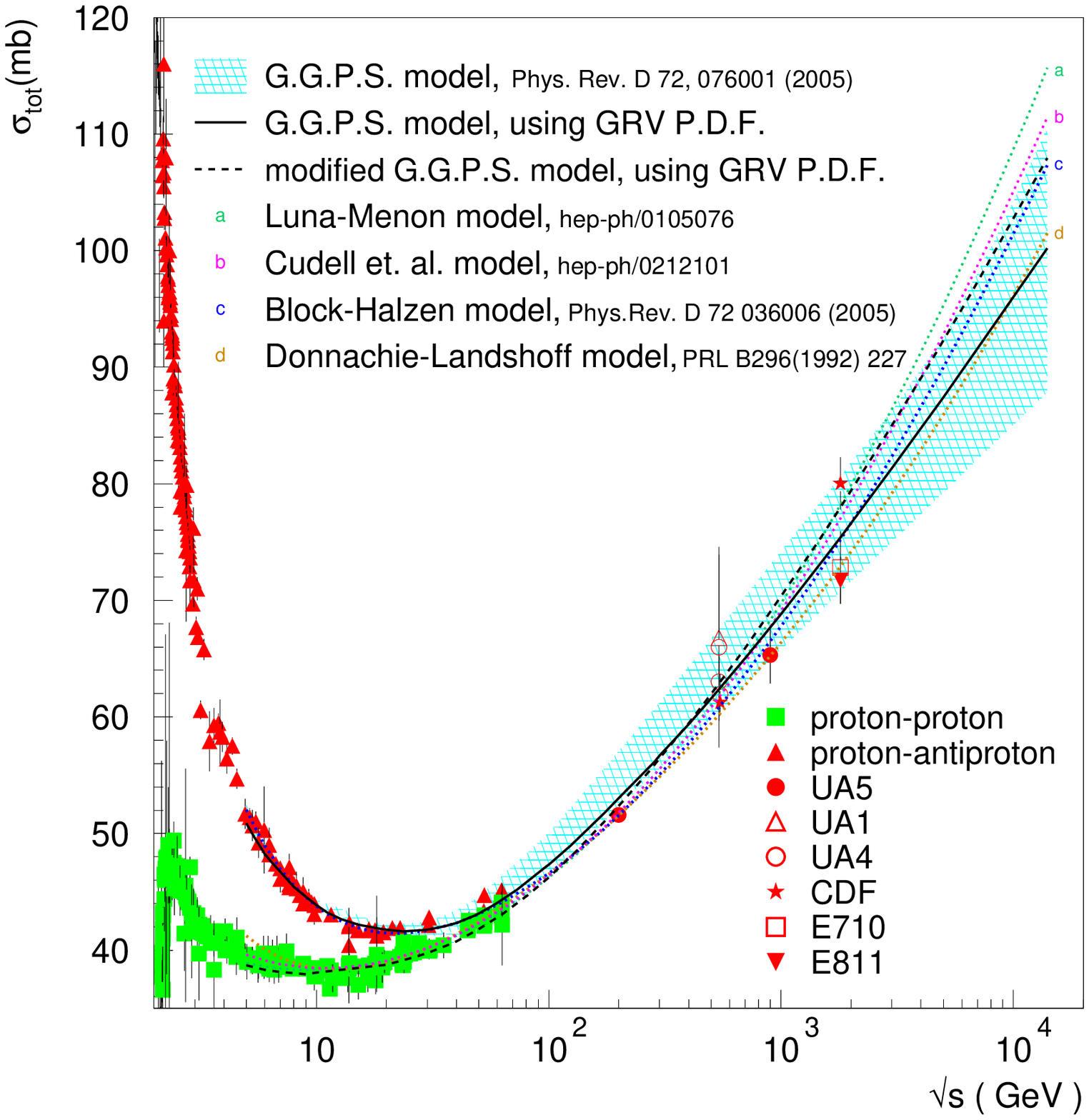}
\caption{Predictions for \protect\sigtot\ in various models. The
shaded area gives the range of results in the eikonalised mini-jet
model with soft gluon re-summation~\protect\cite{Godbole:2004kx} (the
G.G.P.S. model) the solid line giving the prediction obtained using
the GRV parton densities~\protect\cite{Gluck:1991ng} in the model.
Curve $d$ indicates predictions for the
DL fit~\protect\cite{Donnachie:1992ny}. Curve $c$
and the uppermost curve $a$, are the results of two
analytical models incorporating constraints from unitarity and
analyticity, from~\protect\cite{Block:2005ka} and
\protect\cite{Avila:2002tk}, respectively. The prediction obtained
by Igi and Ishida, using FESR follows very closely to that given by the BH
curve. Further, curve $b$ is the result of a
fit by the COMPETE collaboration~\protect\cite{Cudell:2002xe}.
\label{csection} }
\end{figure}

Figure \ref{csection}  summarises the predictions of the different models
described in the previous section.  The shaded area gives
the range of predictions in the eikonalised mini-jet model with soft gluon
re-summation~\protect\cite{Godbole:2004kx} (the G.G.P.S. model), the different
PDF's used giving the range as described in the earlier section.
The solid line gives prediction obtained using the GRV parton
densities~\protect\cite{Gluck:1991ng} in the model. The
curve $d$ indicates predictions of the DL
fit~\protect\cite{Donnachie:1992ny}.  The (BH) curve $c$ and
the uppermost curve $a$, are the results of two analytical models
incorporating constraints from unitarity and analyticity,
from~\protect\cite{Block:2005ka} and \protect\cite{Avila:2002tk}, respectively.
The predictions obtained by Igi and Ishida, using FESR follows
very closely to that given by the BH curve. Further, curve
$b$ is the result of a fit by the COMPETE
collaboration~\protect\cite{Cudell:2002xe}.
The parametrisation for the DL curve and BH curve is already given in the
last section.
It is gratifying to see that the range of results of our QCD motivated
mini-jet models for the LHC span the other predictions obtained in models using
unitarity, factorisation, analyticity along with fits to  the current data.
Thus the two sets of predictions seem consistent with each other.

We have parametrised the results of our EMM model
with a $\ln^2 (s)$ fit. We found that in most cases this gave a better
representation of our results than a fit of the Regge--Pomeron type
of the form of eq. (\ref{DL}).  The top edge of the EMM model prediction is
obtained for the MRST parametrisation whereas the lower edge for the GRV98lo.
We give  fits to our results for $\sigma^{pp}$ of the form
\be
\label{emmfits}
\sigtot = a_0 + a_1 s^{b}  + a_2 \ln (s) + a_3 \ln^2 (s).
\ee
The values of various parameters for the top end lower edge as well as the
central curve are given in table 2.

\begin{table}[t!]
\caption{Values of  $a_0,a_1,a_2,a_3$ and $b$ parton densities in
the proton, for  which the EMM (as described in
ref.~\protect\cite{Godbole:2004kx}) gives a satisfactory description
of \sigtot.}
\vspace{0.5em}

\hskip4pc\vbox{
\columnwidth=26pc
\begin{tabular}{lccccc}
\\[-2mm]
         & $a_0$ (mb)& $a_1$ (mb) & $b$ & $a_2$ (mb)& $a_3$ (mb)
         \\[1mm]
\hline\\[-2mm]
Top edge & \phantom{$-$0}23.61 & \phantom{0}54.62 & $-$0.52 & \phantom{0}1.15 & \phantom{$-$}0.17 \\
Center   & $-$139.80 & 193.89 & $-$0.11 & 13.98 & $-$0.14\\
Lower edge & \phantom{0}$-$68.73&125.80&$-$0.16&11.05&$-$0.16\\[1mm]
\end{tabular}}
\label{param}
\end{table}

\section{Conclusions}

We thus see that the range of the results for the \sigtot\ from our
QCD motivated EMM model~\cite{Godbole:2004kx} spans the range of
predictions made using the current data and general arguments of
unitarity and/or factorisation. Further, we give $\ln^2 (s)$
parametrisation of the model results for \sigtot\  which may be used
in evaluating the  range of predictions for the underlying event
at the LHC.

\section*{Acknowledgements}

This work was supported in part through EU RTN contract no.
CT 2002-0311. RG acknowledges partial support from the
Department of Science and Technology, India
under project no. SP/S2/K-01/2000-II.
AG acknowledges support from MCYT under project no.
FPA 2003-09298-CO2-01.

\end{document}